# Pressure-induced insulator-metal transition in two-dimensional Mott insulator NiPS$_3$

Takahiro Matsuoka[1]*, Amanda Haglund[1], Rui Xue[2], Jesse S. Smith[3], Maik Lang[4], Antonio M. dos Santos[5], and David Mandrus[1,2]

[1]*Materials Science and Engineering, The University of Tennessee, Knoxville, TN 37996, USA*
[2]*Department of Physics and Astronomy, The University of Tennessee, Knoxville, TN 37996, USA*
[3]*HPCAT, X-ray Science Division, Argonne National Laboratory, Argonne, IL 60439, USA*
[4]*Department of Nuclear Engineering, University of Tennessee, Knoxville, TN 37996, USA*
[5]*Neutron Scattering Division, Oak Ridge National Laboratory, Oak Ridge, TN 37831, USA*

The pressure-induced insulator to metal transition (IMT) of layered magnetic nickel phosphorous tri-sulfide NiPS$_3$ was studied *in-situ* under quasi-uniaxial conditions by means of electrical resistance (*R*) and X-ray diffraction (XRD) measurements. This sluggish transition is shown to occur at 35 GPa. Transport measurements show no evidence of superconductivity to the lowest measured temperature (~ 2 K). The structure results presented here differ from earlier *in-situ* work that subjected the sample to a different pressure state, suggesting that in NiPS$_3$ the phase stability fields are highly dependent on strain. It is suggested that careful control of the strain is essential when studying the electronic and magnetic properties of layered van der Waals solids.

## 1. Introduction

The layered transition metal trisulfides ($M$PS$_3$; $M$ = V, Mn, Fe, Co, Ni, etc.) commonly crystallize in a monoclinic structure with $C2/m$ symmetry at ambient pressure.[1–6] The transition-metal ions form a honeycomb lattice, and each layer is isolated between two layers of sulfur and phosphorus triangles (PS$_3$), which in turn are separated by van der Waals (vdW) gaps. Many of these materials are Mott or charge-transfer insulators and show antiferromagnetic (AFM) ordering.[1,2,7–9] The tuning of $M$PS$_3$, and their closely related selenium analogs ($M$PSe$_3$), from an antiferromagnetic Mott insulating state into a metallic, or even superconducting, state is of great interest for fundamental magnetism and Mott physics research. Their superconducting state would strengthen our understanding of the underlying physics for superconducting cuprates, which also arises from a doped antiferromagnetic Mott insulator. Moreover, their tunable electronic properties make those compounds fascinating for potential applications to magnetic materials and spintronic devices.[10]

Pressure is an effective tool to perturb the electronic states of Mott insulators by bandwidth modulation. Pressure-induced insulator-to-metal transition (IMT) has been reported for several materials.[8,11–16] Prior experimental work suggests that the IMTs of those $M$PS$_3$ and $M$PSe$_3$ occur concomitantly with or following structural transformations and spin-crossovers (high-spin to low-spin state transition).[11–13,15,17] Remarkably, FePSe$_3$ exhibits a superconducting transition at 2.5 K and 9 GPa (increasing to 5.5 K at 30 GPa).[8]

More pertinent to the present work, NiPS$_3$ has filled $t_{2g}$ and half-filled $e_g$ orbitals ($|S_Z|$ = 1). Below the Néel temperature ($T_N$) of 155 K, NiPS$_3$ orders magnetically with a propagation vector of $\boldsymbol{k}$ = [010] with the moment direction found to be mostly along the $c$ axis.[3] The half-filled Ni $e_g$ orbitals point towards the S atoms, lying in the out-of-plane direction, and make NiPS$_3$ susceptible to $p$ (S) – $d$ (Ni) hybridization, and thus sensitive to the inter-layer distance.[18] There is a striking prediction that the Mott IMT might be accompanied by the $|S_Z|$ = 1 to $|S_Z|$ = 1/2 state, without changing its crystal structure.[18] Such electronically-driven IMTs are less common among vdW materials. The Mott phase and superconductivity in twisted bilayer graphene and the suggested IMT in the high-pressure phase of V$_{0.9}$PS$_3$ are the only examples to our knowledge.[12,19,20] Also, the absence of structural phase transition suggests that NiPS$_3$ is a promising material for ultrafast resistivity switching behavior applied to thin electronic and spintronic applications.

In the following, we present a brief review of prior high-pressure experiments of NiPS$_3$. Raman scattering measurements on single crystals, using argon (Ar) as a pressure transmitting medium (PTM), suggested that no structural transition occurred up to 27 GPa at room temperature.[21] On the other hand, XRD measurements of a powder sample using silicone-oil (Si-oil) as the PTM reported two structural transformations, first to an intermediate phase at 15 GPa followed by a transition to the high-pressure phase at 27 GPa.[11] Aggregating the data from measurements of powder XRD with Si-oil as the PTM, Raman scattering of a single crystal in KBr at 5 K, and the temperature dependence of the electrical resistance of a single crystal compressed in KBr, allows the mapping of a tentative phase diagram suggesting that the IMT occurs at 20 GPa inside the intermediate phase region.[11] Here, we notice that, owing to the different results arising from the variety of hydrostatic conditions, the question regarding whether the material undergoes an electronic bandwidth-controlled IMT remains unanswered. In addition, the superconducting transition is also still a subject to be explored. This work aims at clarifying the high-pressure phase diagram of $NiPS_3$ and its dependence on the strain state, as these remain open questions.

The technical difficulty in investigating the high-pressure properties of vdW compounds is that pressure, and thus strain, can easily induce inter-layer slipping, which often drives structural transformations. In this study, we revisited the problem by the electrical resistance ($R$) and X-ray diffraction (XRD) measurements of a single crystal, while ensuring the pressure conditions of all experiments were kept as similar as possible. Since we expected $NiPS_3$ to be sensitive to inter-layer compression, we employed quasi-uniaxial pressure perpendicular to the $a-b$ plane (out-of-plane direction) in all measurements. It is noted that we define the quasi-uniaxial pressure as the condition that the pressure along an axis is far larger than along the other two.

2. **Experimental Method**

Synthesis of the $NiPS_3$ crystals was performed through chemical vapor transport (CVT). First, 6 g total of Ni (99.996%, Alfa Aesar), P (99.999+%, Alfa Aesar), and S (99.9995%, Alfa Aesar) in stoichiometric ratio were measured and then ground together with a mortar and pestle, under an Ar-gas atmosphere in a glove box. The homogenized powder was transferred into a die, pressed into a pellet, and the pellet was sealed in a fused quartz tube under vacuum. After annealing for 7 days in a box furnace at 700°C, the resulting grey polycrystalline $NiPS_3$ powder was removed from the tube and verified for phase purity with XRD on a Bruker D2 Phaser. For the CVT growth process, 2 g of $NiPS_3$ powder and 10−15 mg of iodine crystals were sealed under vacuum in a fused quartz tube with an inner diameter of 19 mm and length of around 10 cm. The tube was then placed in a horizontal tube furnace under a temperature gradient of 600−700°C, for 1−2 weeks, to produce the final bulk single crystals. The crystal structure was confirmed through XRD, stoichiometry through energy-dispersive X-ray spectroscopy, and magnetic ordering temperature with SQUID measurements.

Diamond anvil cells (DACs) equipped with type-Ia diamond anvils were used for the high-pressure generation. Figure 1 shows the experimental setup for the electrical resistance measurements. The culet diameters of the diamond anvils were from 300 to 520 μm. The 0.25 mm-thick metal gaskets (stainless steel, SUS310) were pre-indented to 50 μm, and a 100 − 200 μm diameter hole (sample chamber) was drilled in the center of the pre-indented area. A layer of compressed diamond-powder and epoxy resin mixture insulates the electrical leads (platinum (Pt) film, 5 μm in thickness) from the metal gasket. A small single crystal flake of about 10 μm in thickness was exfoliated from a large single crystal. Then, the flake was shaped to match a sample chamber closely. The single crystals were loaded with the *a−b* plane laying on the diamond's flat surface. A pre-compressed sodium chloride (NaCl) flake was placed underneath the sample, serving as a PTM and ensuring that the pressure distribution across the sample was as homogeneous as possible. Several tiny ruby chips (Cr:$Al_2O_3$) were added together with the samples to determine the pressure using the ruby fluorescence method by applying a standard calibration curve.[22] The configuration of the sample and gasket is identical in all experiments. However, we used different DACs, as appropriate for the experimental technique used for each sample. We measured the electrical resistance using a source meter (KEITHLEY 2450) and a nano-voltmeter (KEITHLEY 2182A). The applied current was 0.1 mA in all measurements. The temperature control down to 2 K was achieved by a Physical Property Measurement System (PPMS, Quantum Design). The indicated pressures for electrical resistance were that obtained by averaging the values obtained at room temperature before and after each cooling and warming cycle. The actual pressures at low temperatures are thought to be somewhat different from the indicated values. The reproducibility of the experimental results was ensured by performing three independent electrical resistance measurements (samples No. 1, No. 2, and No. 3). The electrical resistivity ($\rho$) was calculated using the area/length, ($w \times t$) / $l$, ratio, (where *w* is the width of the sample, *t* is the thickness, and *l* is the separation between electrical leads) measured before applying pressure. The $l \times w \times t$ for each sample is 60 μm × 60 μm ×5 μm for No. 1, 40 μm × 50 μm ×5 μm for No. 2, and 100 μm × 40 μm ×5 μm for No. 3. To identify the phases present, we measured XRD for samples No. 2 and 3 at 24 GPa and 28 GPa, respectively. In addition, we measured the XRD of another sample (sample No. 4) to check the crystal structure at low pressure (0.7 GPa). XRD measurements were performed at 16-ID-B at the Advanced Photon

Source (APS), Argonne National Laboratory. The wavelength of the X-ray beam was 0.4066 Å, and the beam spot size was 10 μm in diameter in all measurements. Diffraction data were collected on a flat panel detector array (Dectris Pilatus 1M-F, pixel size: 172 μm × 172 μm) in the forward scattering geometry at room temperature. For two-dimensional XRD data reduction, we used Dioptas.[23] To maximize the number of reflections from the single crystal samples, diffraction measurements were taken with the cell rotated about the vertical axis (± 10 degrees).

## 3. Results and Discussions

Figure 2 shows microscope images of $NiPS_3$ (No. 1−3) as a function of pressure. In all samples, reflectivity increases with pressure. In sample No. 1, it becomes comparable to that of the Pt (electrical leads) at pressures above 31 GPa. The whole area becomes metallic above 38 GPa. In No. 2, the entire sample already appears shiny at 28 GPa. Upon the pressure unloading, the metallic appearance persists down to 14 GPa and gradually vanishes until it completely disappears below 5.8 GPa (See the pictures of Sample No. 2).[24] In No. 3, we did not observe the metallic reflection up to 24 GPa.

Figure 3a shows the pressure dependence of $R$ for samples No. 1−3 at room temperature. All samples exceed the range measurable by our equipment (> 2 MΩ) until approximately 10 GPa. Above this pressure, $R$ decreases, showing a smoothly varying $R$ vs. $P$ slopes up to each sample's highest pressure. Figure 3b shows the temperature dependence of $R$ for sample No. 1 at different pressures. All the resistance curves are normalized to their 250 K value to compare their respective rate of change ($dR/dT$). The $dR/dT$ is negative below 28 GPa, characteristic of semiconductor behavior. The negative slope changes gradually and becomes positive at temperatures above 50 K when the pressure is 31 GPa. At 34 GPa, the $dR/dT$ is positive but shows an upturn below 20 K, suggesting that $NiPS_3$ is semimetallic or the mixed phase of a metal and a semiconductor. At 37 and 40 GPa, the sample exhibits a metallic character with a positive $dR/dT$ in the whole temperature range and an almost constant $R$ at the lowest temperatures. In the present report, we define the IMT pressure as the pressure where the $R$ vs. $T$ curve shows metallic behavior in the whole temperature range. For the sample No. 1, the IMT is 35 GPa. Throughout the measured pressures (in both the semiconducting and metallic states) and across the entire temperature range down to 2 K, there is no evidence of superconductivity, such as an abrupt drop in resistance.

The inset graph of Fig. 3a shows the pressure dependence of the $\rho$. Because visible observation indicated that $w$ and $l$ did not change, we assumed that only the thickness changes. Since the thickness reduces with pressure, the actual $\rho$ should be smaller than the indicated values in the graph. In addition, it is noted that we measured No. 2 by two-point contact method that the two leads contacted the sample and were divided into four in the vicinity of the sample. Thus, the indicated $\rho$ of No. 2 is even bigger than the actual value because it includes the resistance of the electrical leads and the contact resistance between the sample and the leads. The $\rho$ of sample No. 1 is 2.25 mΩ·cm at the IMT (35 GPa) and decreases to 0.88 mΩ·cm in the metallic state at 42 GPa. In sample No.2 at 28 GPa, where the sample appears shiny metallic, the $\rho$ becomes 1.01 mΩ·cm comparable to that of No. 1 in the metallic state, indicating that No. 2 is in the metallic state. For sample No. 3 at 24 GPa, the $\rho$ is 40.0 mΩ·cm showing that the sample is a semiconductor. We note that we did not measure the $R$ vs. $T$ in detail for samples No. 2 and 3 as DACs optimized for diffraction are not compatible with any of our cryostats. However, by cooling DACs in liquid-$N_2$, we confirmed that sample No. 2 and No. 3 showed a negative and a positive d$R$/d$T$ at 24 GPa and 28 GPa, respectively.

  The $R$ of samples No. 2 and 3 was carefully monitored while downloading the pressure (Fig. 3a). As pressure decreases, the $R$ of sample No. 2 increases gradually, showing a large hysteresis against the pressure loading and exhibiting a significant increase below 9 GPa. The decrease in conductivity is consistent with the reflectivity observation in the microscope images (Fig. 2). At 0.2 GPa, the $R$ was above the measurement range of our setup (> 2 MΩ). The decrease in conductivity below 9 GPa occurs in sample No. 3, which was in the semiconductor phase at 24 GPa. The observation of hysteresis will be discussed further in the context of the bonding nature of $M$PS$_3$.

  Fig. 4 shows the XRD profiles for sample No. 4 (0.7 GPa), No. 3 (24 GPa), and No. 2 (28 GPa). We note that the diffraction signal from the sample appears as spot reflections characteristic of scattering from single crystals. Within the area where the X-ray was hitting, the XRD remained single crystal-like upon increasing pressure, with spots showing slight broadening following the structure changes discussed later (See Supplemental materials Fig. S1 for the XRD images).[24] At 0.7 GPa, all the observed XRD peaks of NiPS$_3$ are consistent with the reported $C2/m$ structure (Fig. 4c). The calculated lattice parameters for NiPS$_3$ are $a$ = 5.800(1) Å, $b$ = 10.057(1) Å, $c$ = 6.605(2) Å, and $\beta$ = 106.99(2) deg. in agreement with the reported values.[25] By observing low-angle diffraction near $2\Theta$ = 4.9 deg, it is clear that the high-pressure phases (24 and 28 GPa) have different structures from $C2/m$. The XRD profiles of No. 2 and No. 3 are somewhat similar but have an apparent difference: No. 3 has two peaks at $2\Theta$ = 10.2 and 20.4 deg., while these are absent in No. 2. On the other hand, the intense peak at $2\Theta = 22.7$ deg. in No. 2 is not observed in No. 3. Although we have not identified crystal structures due to the limited number of peaks, our analysis suggests that trigonal systems can be the candidates for each of No. 3 (24 GPa, $a = b = 5.49(2)$ Å, $c = 5.59(3)$ Å, $\alpha = \beta = 90$ deg., and $\gamma = 120$ deg.) and No. 2 (28 GPa, $a = b = 5.47(1)$ Å, $c = 10.87(9)$ Å, $\alpha = \beta = 90$ deg., and $\gamma = 120$ deg.). From the XRD measurements, we conclude that NiPS$_3$ has different crystal structures at 0.7, 24, and 28 GPa. Here, it is clear that NiPS$_3$ exhibits the IMT accompanied by a structural transformation. A Mott transition is normally the first order associated with the change in the cell volume and the electrical resistance. The question of whether NiPS$_3$ exhibits a unit-cell volume change when the IMT occurs remains for future detailed studies.

The obtained XRD profile for the high-pressure phases did not match the data previously reported by the powder XRD measurement performed in a quasi-hydrostatic condition using Si-oil as the PTM (See Fig. S2 in the Supplemental Information for the comparison of profiles).[24] In Si-oil, NiPS$_3$ is in $C2/m$ at 15−30 GPa.[11] The pressure range 27−30 GPa belongs to the transition interval to a higher-pressure phase predicted to adopt a $P\bar{3}m$ structure.[11] The observed difference is likely to originate from the different PTM leading to different strain states on the sample. In this study, we applied quasi-uniaxial pressure perpendicular to the $a$–$b$ plane of NiPS$_3$. In the powder XRD, crystals randomly orient in the fluid at the beginning of the compression. When Si-oil solidifies (above 5 GPa), the Si-oil develops quasi-uniaxial stress as pressure increases.[26] Since crystals are still randomly oriented in solid Si-oil, the strain conditions become different between crystals, depending on the orientation in the sample chamber. Crystals with $a$–$b$ planes aligned along the pressurizing direction would be exposed to bigger shear stress. Also, some others experience a preferred compression of the $c$-axis (See Supplemental Information Fig. S3 for a schematic description).[24] These observations suggest that the pressure condition significantly affects the high-pressure phase diagram of NiPS$_3$.

Below, we discuss the transition process from an insulator to a metal. By comparing the data of this study and previous reports, we notice the IMT pressure varies in the samples. Sample No. 1 completes the IMT above 34 GPa, while No. 2 is metallic at 28 GPa (Fig. 2 and Fig. 3a). The previous study reported the IMT at 15 GPa for a single crystal.[11] Considering the weak interlayer vdW interaction, it is straightforward to think that the interlayer slipping, hence the shortening of inter-atomic distance and the transition to the metallic phase, can be promoted by shear stress at lower pressures. The strain environment realized in each measurement cannot be exactly the same even though a similar experimental setup is employed. For example, when the surfaces of diamond anvils are off-parallel during compression, shear stress emerges on the sample. Clearly, the diamond anvils cannot be ensured to be perfectly parallel in every experiment, especially at high pressure. When different PTMs are used, strain environments are inevitably different. From our observations here, we postulate that sample No. 2 was exposed to larger shear stress than No. 1. Although the details of experimental setups in the measurement in Ref. 11 are unknown, we speculate that their sample had higher shear stress than that we examined. Another important point is that $NiPS_3$ is a fragile sample that needs to be cut to fit the DAC sample chamber, thus, as another possibility, the location where the sample is harvested from and subsequently cut and its original thickness can also affect the initial strain state, making them more prone to shear under pressure. From another perspective, it would be interesting to study the electrical and magnetic characteristics of $NiPS_3$ by applying a more controlled strain such as shearing and twisting, as it might be possible to obtain a metallic phase at much lower pressures.

Comparing the high-pressure properties of $NiPS_3$ and $FePS_3$, we notice an apparent difference. The $R$ vs. $T$ curve of $NiPS_3$ shows a large hysteresis between pressure loading and unloading, while $FePS_3$ does not.[8] Excluding from the following considerations the possibility of a wildly different strain state for the $FePS_3$ sample in Ref 8, and although the crystal structures of the high-pressure phases of $NiPS_3$ are yet to be determined, it is reasonable to assume that the bonding nature of the two materials is different. In $MPS_3$, the $t_{2g}$ and $e_g$ orbitals play roles in bonding. The hopping integrals between $t_{2g}$ and $e_g$ show different anisotropy; the $t_{2g}$ and $e_g$ yield strong in-plane $d-d$ and inter-plane $d-p-p-d$ overlap, respectively.[18] In $FePS_3$, the $t_{2g}$ and $e_g$ orbitals are partially filled at ambient pressure. In $NiPS_3$, only the $e_g$ orbitals are half-filled ($t_{2g}$ are filled). Therefore, it is expected that the crystal structures and corresponding electronic band structures of $FePS_3$ and $NiPS_3$ show different responses to compression.

## 4. Conclusions

We studied the electrical transport and structure of $NiPS_3$ under quasi-uniaxial pressure in the layer-stacking direction. The electrical resistance significantly and continuously decreases as the pressure increases, consistent with the increase of the optical reflectivity of the sample. $NiPS_3$ is an insulator at ambient pressure and becomes a metal above 35 GPa. The XRD measurements confirmed that the IMT is accompanied by a structural change. Metallic $NiPS_3$ shows no superconducting transition in the measured temperature range, down to 2 K. The IMT pressure varies depending on the hydrostaticity of the pressure loading and thus the strain environment. As observed for NiPS$_3$, it is expected that hydrostaticity is the key factor to determine the stability fields of the high-pressure phases of $M$PS$_3$ and $M$PSe$_3$. It would be worth investigating their structural and electronic transitions while varying the strain state under carefully controlled conditions.

**Acknowledgments**

Acknowledgment follows the main text of the paper. No section numbers should be given. This research is funded by the Gordon and Betty Moore Foundation's EPiQS Initiative, Grant GBMF9069 to D.M. M.L. acknowledges support by the U.S. Department of Energy (DOE), Office of Science, Basic Energy Sciences (BES), under Award No. DE-SC0020321. A portion of this research used resources at the Spallation Neutron Source, a DOE Office of Science User Facility operated by the Oak Ridge National Laboratory. XRD experiments in this study were performed at HPCAT (Sector 16), Advanced Photon Source (APS), Argonne National Laboratory. HPCAT operations are supported by DOE-NNSA's Office of Experimental Sciences. The Advanced Photon Source is a U.S. Department of Energy (DOE) Office of Science User Facility operated for the DOE Office of Science by Argonne National Laboratory under Contract No. DE-AC02-06CH11357. A part of HPCAT beam time was provided by the Chicago/DOE Alliance Center. We appreciate valuable and fruitful discussions by Prof. Janice Musfeldt, Prof. Heung-Sik Kim, Dr. Subhasis Sanmanta, and Mr. Nathan Cassidy Harms.

*E-mail: tmatsuok@utk.edu

**Figure Captions**

Fig. 1. (color online) The experimental setup for the electrical resistance and XRD measurements in the DAC. (a) The schematic cross-section of the DAC. A small ruby chip is also added to the sample chamber (not pictured). The X-ray beam hits the position in-between the electrical leads. (b) Sample No. 1 at 6.6 GPa, looking through a diamond anvil. The dashed circle indicates the position of a ruby chip.

Fig. 2. (color online) $NiPS_3$ under pressure and room temperature (296 K) in the three independent measurements (Samples No. 1−3). 'S' indicates the samples. The pictures in a dashed square were collected while pressure unloading. The light intensity, exposure time, and gain of the microscope camera were kept constant throughout. In the measurement of sample No. 2, there were a couple of extra electrical leads, shown at 3 and 7 o'clock in the pictures. They were not used for the measurements.

Fig. 3. (color online) Electrical resistance measurements of $NiPS_3$ under high pressure. (a) Pressure dependence of the $R$. The data obtained in three independent measurements are plotted together with Ma *et al*.[11] The solid vertical arrows are the IMT pressures of sample No. 1 (red) and Ma *et al*. [11] The inset is the pressure dependence of the $\rho$. (b) Temperature dependence of the $R$ for sample No. 1. The data are normalized to their value at 250 K.

Fig. 4. (color online) The XRD profiles of $NiPS_3$ samples No. 2 (28 GPa), No. 3 (24 GPa), and No. 4 (0.7 GPa) at room temperature. All profiles are background subtracted. The profile of sample No. 3 in the low-angle region (4.5–5.5 deg.) is vertically expanded (10 times) for the visibility of a small peak near 4.9 deg. The region $2\Theta$ = 5.5–7.5 deg. and 12–13.8 deg. are excluded from the profiles because we did not observe peaks except for NaCl and Pt.[27–29] NaCl exhibits the B1 (NaCl-type) → B2 (CsCl-type) structure transformation near 30 GPa.[27,28]

Fig. 1

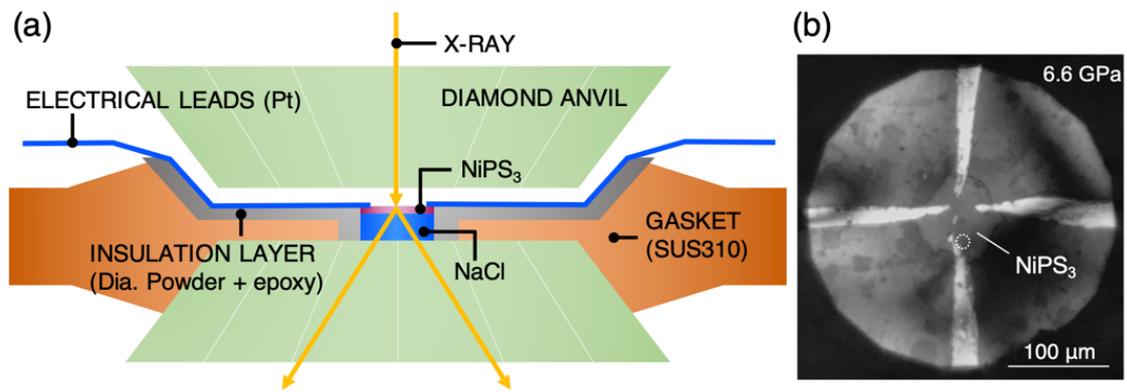

Fig. 2

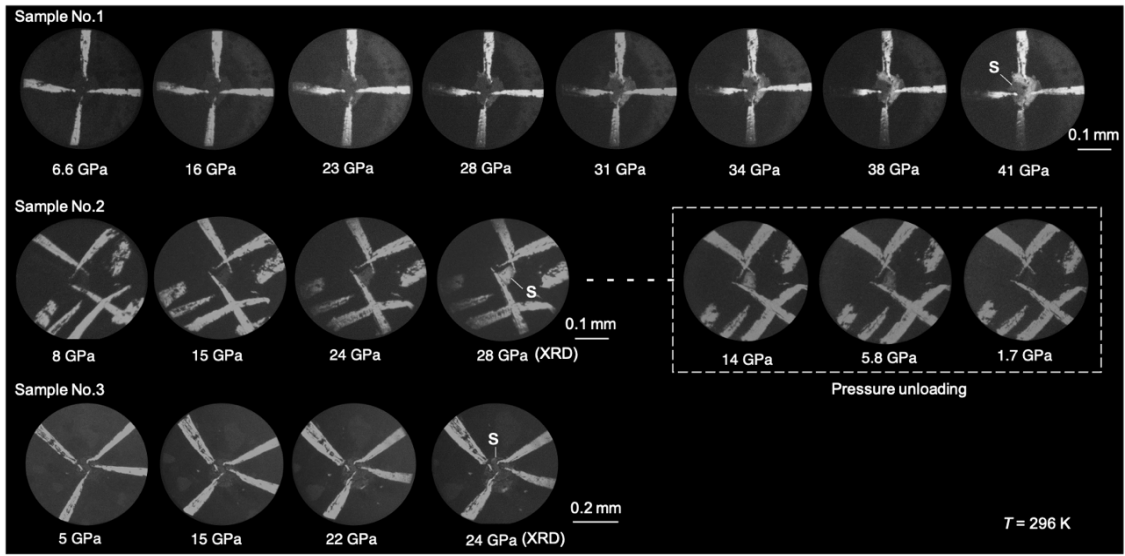

Fig. 3

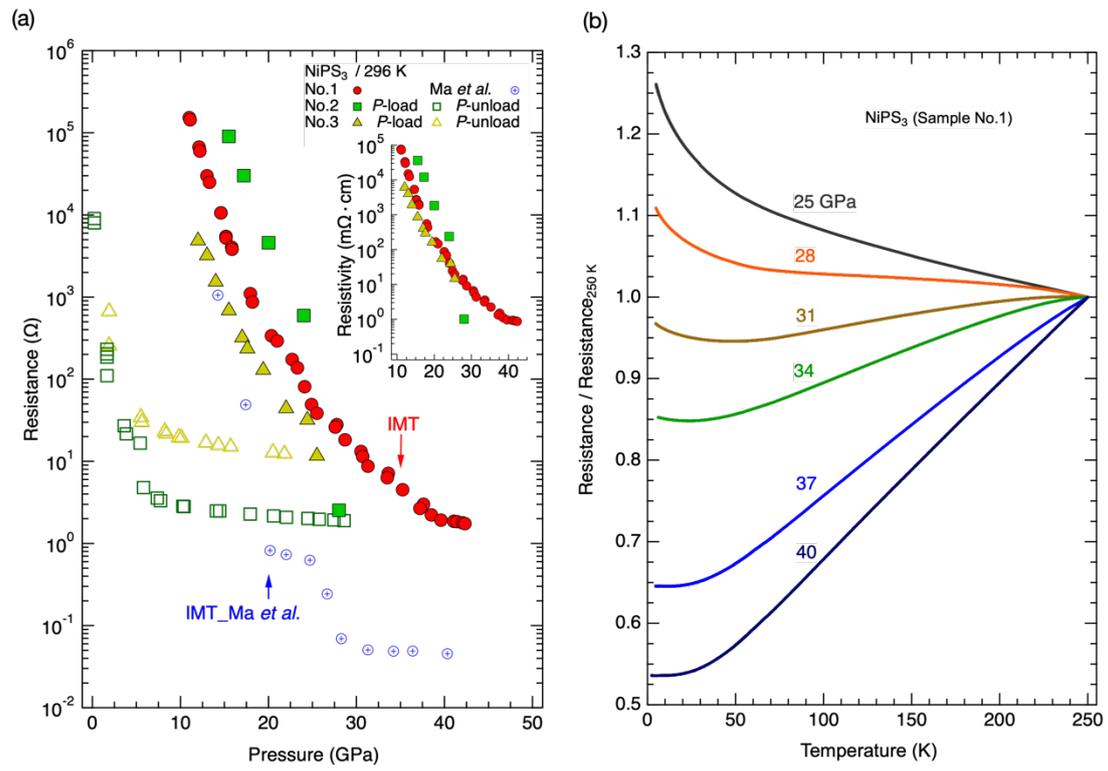

Fig. 4.

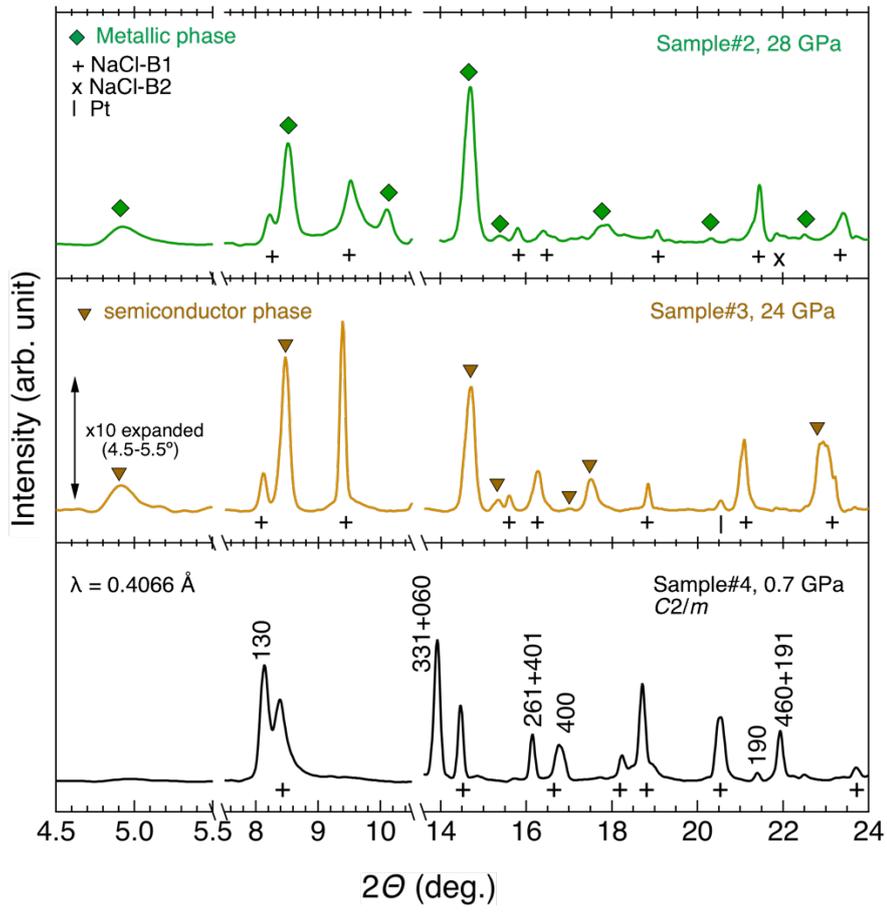